\documentclass[pra,twocolumn,showpacs,superscriptaddress]{revtex4}
\usepackage[dvips]{graphicx}
\usepackage{amsmath,amsfonts,amssymb}
\usepackage{dcolumn}
\usepackage{bm}
\usepackage{epsfig}
\def\jpb{J. Phys. B: At. Mol. Opt. Phys.}
\def\etal{{\em et al.}}

\def\beq{\begin{equation}}
\def\eeq{\end{equation}}

\newcommand{\vecE}{{\bf E}}

\newcommand{\vecr}{{\bf r}}

\newcommand{\vecx}{{\bf X}}
\newcommand{\vecf}{{\bf f}}
\newcommand{\vecF}{{\bf F}}
\newcommand{\vecy}{{\bf y}}

\begin{document}
\title{Harmonic emission from cluster nanoplasmas subject to intense short laser pulses}
\date{\today}
\author{S.V.\ Popruzhenko}
\affiliation{Max-Planck-Institut f\"ur Kernphysik, Postfach 103980,
69029 Heidelberg, Germany}
\affiliation{Moscow State Engineering Physics Institute, Kashirskoe Shosse 31, 115409, Moscow, Russia}
\author{M.\ Kundu}
\affiliation{Max-Planck-Institut f\"ur Kernphysik, Postfach 103980,
69029 Heidelberg, Germany}
\author{D.F.\ Zaretsky}
\affiliation{Max-Born-Institut, Max-Born-Str. 2A, 12489 Berlin, Germany}
\affiliation{Russian Research Center ``Kurchatov Institute'', 123182 Moscow, Russia}
\author{D.\ Bauer}
\affiliation{Max-Planck-Institut f\"ur Kernphysik, Postfach 103980,
69029 Heidelberg, Germany}

\date{\today}

\begin{abstract}
Harmonic emission from cluster nanoplasmas subject to short intense infrared laser pulses is studied.
In a previous publication [M.\ Kundu {\em et al.}, \pra {\bf 76}, 033201 (2007)] we reported particle-in-cell simulation results showing resonant enhancements of low-order harmonics when the Mie plasma frequency of the ionizing and expanding cluster resonates with the respective harmonic frequency. Simultaneously we found that high-order harmonics were barely present in the spectrum, even at high intensities.
The current paper is focused on the analytical modeling of the process.
We show that dynamical stochasticity owing to nonlinear resonance inhibits the emission of high order harmonics.
\end{abstract}


\maketitle

\section{Introduction}
The study of rare gas and metal clusters interacting with intense infrared, optical, and ultraviolet laser pulses has emerged as a new promising research area 
in strong field physics (see \cite{rost06,krainov02} for reviews).
The generation of fast electrons and ions, the production of high charge states, the generation of x-rays and nuclear fusion (in the case of deuterium-enriched clusters) was observed using  cluster targets in strong laser fields of intensities up to $10^{20}$W/cm$^2$.
Hot and dense nonuniform, nonequilibrium and nonstationary plasmas produced under such conditions and confined on a femto- or picosecond time scale to nanometer sizes---so-called {\it nanoplasmas}---are new physical objects with unusual properties.

One of the most important features of nanoplasmas is the very efficient energy transfer from light to charged particles, which is much higher (per particle) than for an atomic gas of the same average density \cite{ditmire97}.
Although the particular mechanisms responsible for the laser energy deposition in clusters still remain debated \cite{kuhl,last,korn04,milchberg,kost05,mulser05,kundu06,devis,korn07}, the pivotal role of collective plasma dynamics (in particular the excitation of surface plasmons) and of nonlinear resonance (a definition is given below in Sec.~\ref{reoloh}) in the energy transfer from laser light to the electrons in the nanoplasma and the subsequent outer ionization was proved in experiments, simulations, and simple analytical models \cite{korn04,mulser05,kost05,antonsen05,doeppner,kundu06}.

While the energy absorption from intense laser pulses by nanoplasmas has been widely studied both in experi\-ments and theory, much less attention has been paid to harmonic emission from such systems.
Naturally the question arises whether laser-driven clusters, being highly nonlinear systems, may be a source of high-order harmonics as efficient as atoms in the gas phase (or even more efficient?).
There are at least two essentially different physical mechanisms which could be responsible for emission of harmonics from such a system.

First, the mechanism based on recombination of the virtually ionized electron with its parent ion \cite{corkum}, well-studied in gaseous atomic targets, may also work in clusters where it can be modified by the fact that the atoms are closer to each other, so that the electron's motion between ionization and recombination can be distorted by the field of the other ions. Moreover, the electron may recombine with an ion different from its parent.
This leads to modifications in the single-electron dynamics and in the phase matching conditions while the physical origin of harmonic generation (HG) remains the same as in a gas jet.
Modifications of the atomic recombination mechanism in clusters have been considered in \cite{rost00,maquet01}.
However, in high intensity fields where each atom is loosing one or several electrons already during the leading edge of the laser pulse the recombination mechanism can hardly be efficient.

Second, as dense electron plasmas is produced inside a cluster, its coherent motion may be an efficient source of radiation.
In a collisionless plasma, as it is generated by intense laser fields inside small clusters, individual electron-electron and electron-ion collisions cannot destroy the coherency of the collective electron motion. 
This coherent, collective electron motion may cause HG if it is nonlinear.
Recently, several experiments on high harmonic emission from plasma surfaces illuminated by short laser pulses of intensity $10^{17}$W/cm$^2$ and higher were reported (see, e.g.,  \cite{nph06,nph07}).
The physical picture which is behind such {\em plasma harmonics} appears to be more complicated and diverse than the recombination mechanism in atomic HG.

In macroplasmas the magnetic component of the Lorentz force is a typical source of nonlinearity.
In this case the nonlinearity parameter is $v/c$ where $v$ is the typical velocity related to collective oscillations of the electron plasma and $c$ is the speed of light.
As a consequence, generation of harmonics from dense macroplsmas may require relativistic intensities \cite{nph06}.
As compared to macroplasmas, clusters introduce an extra source of nonlinearity due to their small spacial size, namely $X/R_0$, where $R_0$ is the cluster radius and $X$ is the amplitude with which the electron cloud oscillates under the action of the laser field \cite{fom03}.
Therefore, one could expect strongly nonlinear electron motion even in the nonrelativistic regime.
It should be noted, however, that although individual collisions may not be important, there are other effects which can spoil the coherency required for efficient radiation.
Most of these undesirable effects can be attributed to dynamical instabilities induced by the interaction of particles with the mean self-consistent field (in the presence of the laser field).
Therefore, the examination of {\em plasma harmonics} usually requires not only the analysis of the collective motion but that of individual electron trajectories as well.

Up to now only a few experiments on HG from clusters  are known.
In Refs.~\cite{ditmire96,ditmire97a,vozzi05} HG from rare-gas clusters irradiated by infrared pulses of moderate ($\simeq 10^{13}-10^{14}$W/cm$^2$) intensity was studied.
It was shown that under such conditions harmonics can be generated up to higher orders and with a higher saturation intensity than in a gas jet.
In addition other interesting properties including different power laws in the intensity-dependent harmonic yield for particular harmonics for clusters and atoms  were reported \cite{ditmire96}.
However, in Refs.~\cite{ditmire96,ditmire97a,vozzi05} the applied intensities were rather low to create a dense nanoplasma inside clusters and the main features of the recorded spectra (like a plateau followed by a cutoff) were found to be quite similar to the case of a gaseous target.
This allows to attribute the observed effects to HG along the standard atomic recombination mechanism as described above.

Recently the experimental observation of the third harmonic (TH) generation from argon clusters in  a strong laser field was reported \cite{ditmire07}.
The laser intensity was varied between $10^{14}$ and $10^{16}$W/cm$^2$ and was thus sufficiently high for the creation of a dense nanoplasma.
At such intensities HG along the atomic recombination mechanism relevant in gases is essentially suppressed because of saturated single-electron ionization so that the observed TH signal can be fully attributed to the nonlinearity of the laser-driven nanoplasma.
A resonant enhancement of the TH yield (when the Mie frequency of the expanding cluster approaches three times the laser frequency) has been measured using a pump-probe setup. The TH enhancement of the single-cluster response, as studied in theory before \cite{fom03,fomytski04,fom05}, is, however, masked in the experiment by phase matching effects whose optimization at high average atomic density necessary to create clusters has been shown to be more intricate than for rare gas jets.
The latter complication makes an experimental study of nanoplasma radiation a difficult task while in computations it can be simplified by first examining  the single-cluster response and, second, analyzing propagation effects.

In the recent paper \cite{kundu07}, we considered the radiation emitted by a single cluster exposed to a strong laser pulse.
We computed harmonic spectra from argon clusters in short 800-nm pulses of intensity $2.5\times 10^{14}-7.5\times 10^{17}$W/cm$^2$
using a 3D particle-in-cell (PIC) code (applied before to the study of collisionless energy absorption in laser-driven nanoplasmas \cite{kundu06}).
The most intriguing outcome of our study was the absence of high-order harmonics in the computed spectra, even at high intensities for all cluster sizes we considered.
We attributed this effect to the above-mentioned dynamical instability in motion of individual electrons.

In this paper we introduce two analytical models which describe the collective and single-electron dynamics of a laser-driven nanoplasma, respectively,  and apply them for the explanation of both the numerical results \cite{kundu07} and the data \cite{ditmire07}.
We show that the numerical results can be well described and understood within the rigid sphere model (RSM) but only for low harmonic orders $3,5,7$ while for higher harmonics the RSM yields qualitatively wrong predictions.
Using a simple 1D model we describe a stochastic, resonant single-particle electron dynamics which suppresses the emission of high harmonics.

The paper is organized as follows.
In Sec.~\ref{softpapr} we formulate the statement of the problem and describe the numerical method and the PIC results.
In Sec.~\ref{reoloh} the spectra extracted from the PIC simulations are compared with the predictions of the RSM.
In Sec.~\ref{sohhvsboie} we introduce a model for the description of the single-electron dynamics and radiation and use it to explain the suppression of the high harmonic yield. The last section contains the conclusions.

\section{Statement of the problem and previous results} \label{softpapr}
A cluster is converted into a dense electron plasma almost promptly if the laser pulse is intense enough to ionize the cluster constituents.
We refer to this process as inner ionization which should not be confused with outer ionization, when the nanoplasma electrons leave the cluster.
Several competitive processes govern the evolution of this plasma during the interaction with the pulse and later, until the cluster becomes dissolved due to Coulomb explosion or hydrodynamic expansion.
The electron density increases because of further ionization of atoms and ions by the local electric field which may differ essentially from the applied laser field.
With increasing plasma density due to inner ionization, the oscillating electric field of an infrared laser is screened, so that its amplitude inside the cluster may be a few times or even an order of magnitude less than the amplitude of the incident wave (see more explanations on screening in Sec.~\ref{sohhvsboie} below Eq.(\ref{E0})).
Simultaneously, as soon as a sizeable fraction of electrons have left the cluster, a quasistatic space charge field is built up which may become strong enough to induce further inner ionization (``ionization ignition'', \cite{rose97,bauer03}).
On the other hand, both outer ionization and the expansion of the cluster reduce the electron density.
The net result of this competition is very sensitive to all parameters, including laser intensity, pulse duration, cluster size and type of atoms.
However, for the vast majority of parameters a significant part of the electrons remains confined within the expanding ionic core.
During this stage of the cluster evolution until the laser pulse is off the nonlinear motion of the laser-driven nanoplasma may cause emission of laser harmonics.

We have observed the above described scenario in PIC simulations of laser-cluster interaction, as reported in \cite{kundu06,kundu07}.
The dynamics and the radiation of nanoplasmas was studied for ${\rm Ar}_N$ clusters (with the number of atoms $N\approx 10^4$--$10^5$ and radii $R_0\approx 6$--$10$\,nm), irradiated by linearly polarized, 8-cycle sin$^2$-laser pulses with an electric field $\vecE_l(t)=\vecE_0\sin^2(\omega_l t/2n)\cos(\omega_l t)$ and the wavelength $\lambda = 800$~nm.
Here $\omega_l$ is the carrier frequency and $n=8$ is the number of optical cycles in the pulse.
The contribution of electrons remaining bound in atoms (ions) during the interaction was not taken into account.
Results were reported in Ref.~\cite{kundu07} where also more details about the numerical simulation may be found. Here we restrict ourselves to briefly summarize  the main results:
\begin{itemize}
\item[(i)] The relative yields of low order harmonics depend on the laser intensity and the cluster size.
\item[(ii)] The time-frequency analysis of harmonic spectra shows that low order harmonic enhancements occur when multiples of the laser frequency resonate with the transient Mie frequency.
\item[(iii)] Even for the very high intensity $7.5\times 10^{17}$W/cm$^2$ no distinguishable high harmonics (higher than the 7th) appear in the spectra (see Fig.~1 in Ref.~\cite{kundu07}).
\item[(iv)] Only a part of the nanoplasma deeply bound inside the ion core contributes to (low order) harmonic generation.
\end{itemize}

\section{Resonant enhancement of low-order harmonics} \label{reoloh}
\begin{figure}
\includegraphics[width=0.4\textwidth]{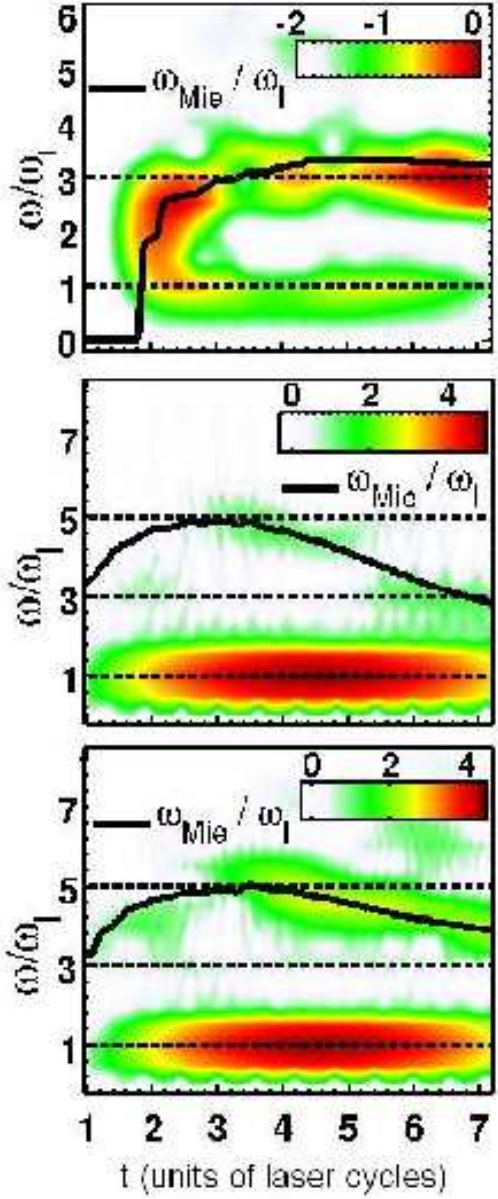}
\caption{(Color online) Time-frequency diagrams  for an ${\rm Ar}_{17256}$ cluster at laser intensities $2.5\times 10^{15}$W/cm$^2$ (a), $7.5\times 10^{17}$W/cm$^2$ (b). Panel (c) shows the TF diagram for an ${\rm Ar}_{92096}$ cluster at the intensity $7.5\times 10^{17}$W/cm$^2$. The scaled time-dependent Mie frequency $\omega_{\rm Mie}(t)/\omega_l$ is included in all plots (solid lines).
\label{fig1}}
\end{figure}

The enhancements of particular low-order harmonics predicted in \cite{fom03,fomytski04} have been later studied numerically for small clusters using molecular dynamics simulations \cite{fom05} and were finally observed in the experiment for the case of the third harmonic \cite{ditmire07}.
It is well established now that the physical origin of enhancements is the resonance between the harmonic frequency and the Mie frequency of the expanding nanoplasma.
The time-frequency analysis of the radiation as calculated from the PIC results for the total acceleration further confirmed this statement \cite{kundu07}.
Typical TF diagrams, showing  which frequencies are emitted when, are reproduced in Fig.~\ref{fig1}.
In the same plots we show the scaled time-dependent effective Mie frequency $\omega_{\rm Mie}(t)/\omega_l$.
The standard definition of the Mie frequency $\omega_{\rm Mie}=\sqrt{4\pi\overline{z}n_0/3}$, where $\overline{z}$ is the average ion charge and $n_0$ is the atom density, is only appropriate for the case of an almost homogeneous charge distribution (in this section atomic units $\hbar=m=e=1$ are used).
However, because of the ignition effect and the cluster expansion the average ion charge $\overline{z}$ depends on the ion position.
As we found in \cite{kundu07}, the low-order harmonics are emitted mainly by electrons confined within the ion core and with  excursion amplitudes comparable to the initial cluster radius $R_0$ or less. In the following we will refer to these electrons as the {\em deeply bound} electrons, and we define the effective Mie frequency as $\omega_{\rm Mie}(t) = \sqrt{Q_0(t)/R_0^3}$ with $Q_0(t)$ the total ionic charge inside the sphere of radius $R_0$ within which the cloud of deeply bound electrons oscillates.

It is clearly seen from the TF diagrams of Fig.\ref{fig1} that the enhancements of the third and fifth harmonic must be attributed to the respective high-order nonlinear resonances between the laser frequency and the Mie frequency, $\omega_{\rm Mie}(t) = m \omega_l$ (with $m=3,5$).
Here we call the resonance {\em nonlinear} since it appears in a {\em nonlinear} system.
This nonlinearity (i.e., anharmonicity in the effective potential of electron-ion interaction) is a necessary condition for the emission of harmonics.
The physical origin of this nonlinearity may be either the electrons jutting out of the core during their motion, thus sensing the Coulomb tail \cite{fom03}, or the inhomogeneous charge distribution within the ion core \cite{fomytski04}.
Note that the appearance of a significant signal at frequencies different from odd multiples of the laser frequency (namely, the second harmonic in Fig.\ref{fig1}(a) and the forth harmonic in Fig.\ref{fig1}(c)) should be attributed to the excitation of eigenoscillations of the electron cloud with the time-dependent eigenfrequency $\omega_{\rm Mie}(t)$.
These oscillations are of significance only in short pulses, as used in the PIC simulations.

The results of the TF analysis in Ref.~\cite{kundu07} not only explain, at least qualitatively, the resonant enhancement of the TH observed in \cite{ditmire07} but also confirm the idea formulated there that with a proper adjustment of parameters harmonics of higher order (5th, maybe 7th) could also be resonantly enhanced.

It was shown in earlier studies \cite{fom03,fom05} that the resonant enhancements of  low-order harmonics may be reasonably described using a very simple rigid sphere model (RSM).
Here we show that, while for low-order harmonics the RSM works well, it fails even qualitatively to reproduce the high-energy part of the spectrum.
In a RSM \cite{parks,mulser05,kundu06} it is usually assumed that both ions and electrons form homogeneous, rigid spheres with sharp boundaries.
In this case the electron and ion charge density distributions are
\beq
\rho_{e(i)}(r)=\mp\overline{z}n_0\theta(R_0-r),
\label{rho0}
\eeq
where $n_0$ is the atom density in the cluster and $\theta(x)$ is the Heaviside step function.
The signs $\mp$ correspond to electrons and ions, respectively.
Within this model the restoring force $\vecF_{\rm ei}\equiv\omega_l^2R_0\vecf$ depends upon the displacement $\vecx\equiv R_0\vecy$ of the electron cloud as \cite{parks}
$$
\vecf(\vecy)=-\frac{\vecy}{y}g_0(y),
\nonumber
$$
\beq
g_0(y)=\left (\frac{\omega_{\rm Mie}}{\omega_l}\right)^2\times
\left\{\begin{array}{l}\displaystyle y-\frac{9y^2}{16}+\frac{y^4}{32},~0\le y\le 2,\\
\displaystyle\frac{1}{y^2},~~~~~~~~~y\ge 2.\end{array}
\right.
\label{g0}
\eeq
Here the dimensionless coordinate $\vecy$ and force $\vecf$ are introduced. The equation of motion reads
\beq
\frac{d^2\vecy}{d\varphi^2}=\vecf(\vecy)-\gamma\frac{d\vecy}{d\varphi}-\frac{\vecE_l(\varphi)}{\omega_l^2R_0}
\label{NE0}
\eeq
where $\vecE_l(\varphi=\omega_l t)$ is the electric field of the laser wave and $\gamma$ is the effective damping constant which can be estimated assuming a collisionless damping mechanism \cite{korn04,korn07} (in calculations we use $\gamma=0.1$, in accordance with such estimates).

\begin{figure*}
\includegraphics[width=0.5\textwidth,angle=-90.0]{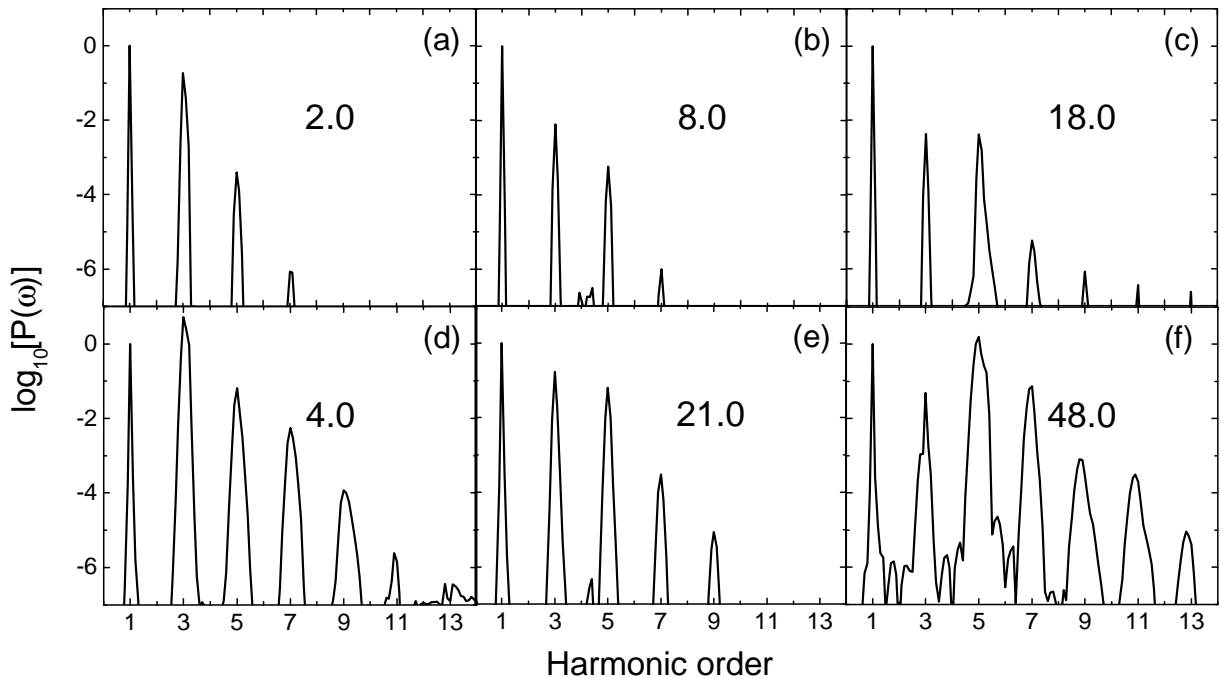}
\caption{Harmonic spectra (normalized to the signal at the fundamental frequency) calculated within the RSM for a cluster of radius $R_0=6.2$nm, $a=R_0/R_e=2$, and ${\rm max}[\omega_{\rm eff}]\approx 3.2\omega_l$ (a,d), ${\rm max}[\omega_{\rm eff}]\approx 4.4\omega_l$ (b,e) and ${\rm max}[\omega_{\rm eff}]\approx 5.4\omega_l$ (c,f). The laser intensities are given in the panels in units of $10^{16}$W/cm$^2$.
\label{fig2}}
\end{figure*}

Using the RSM significant insights into the absorption mechanisms in clusters \cite{kundu06} and TH emission \cite{fom03} have been gained.
However, an essential shortcoming of the RSM in the form specified by Eqs.(\ref{rho0}),(\ref{g0}) is the appearance of even powers of $y$ in (\ref{g0}) at $y\le 2$. Physically it is the result of the discontinuous charge distributions (\ref{rho0}) at $r=R_0$.
As a consequence, in a perturbative treatment of harmonic generation  the intensity of the $s$-th harmonic is proportional to $E_0^{2s-2}$ instead of $E_0^{2s}$.
For example, for the third harmonic one obtains with the RSM $P_{3\omega}\sim E_0^4$, which is obviously unphysical \cite{fom03}.
To improve the model it is sufficient to assume a smoothed charge distribution for the electrons while the positive charge density still may be described by (\ref{rho0}).
We use a Gaussian of characteristic width  $R_e$ for the electron charge density,
\beq
\rho_e(r)=-\overline{z}n_0\exp\left(-r^2/R_e^2\right),
\label{rho1}
\eeq
and assume that the net charge density is zero at the cluster center.
From (\ref{rho1}) the total number of electrons is $N_e=\pi^{3/2}\overline{z}n_0R_e^3$.
The restriction $N_e\le\overline{z}N$ gives  $R_e\le (4/3\sqrt{\pi})^{1/3}R_0\approx 0.91R_0$.
The equation of motion has now the same form (\ref{NE0}) but with a modified restoring force:
$$
g_0(y)\to g_1(y)=\frac{\omega_{\rm Mie}^2}{\omega_l^2 y^2}\times
\nonumber
$$
$$
\times\left\{\frac{1}{2}\bigg\lbrack(y^3+1){\rm erf}[a(y+1)]-(y^3-1){\rm erf}[a(y-1)]\bigg\rbrack+\right.
\nonumber
$$
$$
\left.+\frac{1}{4\sqrt{\pi}a^3}\bigg\lbrack e^{-a^2(y-1)^2}[1-2a^2(1+y+y^2)]-\right.
\nonumber
$$
\beq
\left.-e^{-a^2(y+1)^2}[1-2a^2(1-y+y^2)]\bigg\rbrack\right\}
\label{g1}
\eeq
where $a=R_0/R_e>1.1$ and ${\rm erf}(x)=2/\sqrt{\pi}\int_0^x\exp(-z^2)dz$ is the error function.
Contrary to Eq.~(\ref{g0}), a decomposition of the restoring force (\ref{g1}) contains only odd powers of $y$ while it has the same asymptotic Coulomb behavior for large displacements.
The respective asymptotic expansions have the form [compare with (\ref{g0})]
\beq
g_1(y)=\left (\frac{\omega_{\rm Mie}}{\omega_l}\right)^2\times
\left\{\begin{array}{l}\displaystyle A_1y+A_3y^3+...,~~~~~ y\ll 1,\\
\displaystyle\frac{1}{y^2}+O(e^{-a^2y^2}),~~~~~~~y\gg 1\end{array}
\right.
\label{g1as}
\eeq
with $A_1={\rm erf}(a)-2a/\sqrt{\pi}\exp(-a^2)>0$ and $A_3=-4a^5/5\sqrt{\pi}\exp(-a^2)$. 
It is clear that $A_1\to 1$ if $a\to\infty$, i.e., when the tail of the electron density distribution does not stick out of the ion core.
In this case the eigenfrequency of small oscillations $\sqrt{A_1}\omega_{\rm Mie}$ is equal to $\omega_{\rm Mie}$.
For a finite $a$ the spread of the electron cloud beyond the ion core reduces the eigenfrequency.

Note that within the model we describe above the restoring force is nonlinear because a part of the electron cloud spreads out of the ion core.
In this case the leading term in the nonlinear part of the force can be estimated as $\omega_{\rm Mie}^2\vecx X^2/R_0R_e$.
As a result, the third harmonic signal scales as $R_0^4$ with the cluster radius, in contrast to the standard $R_0^6$ low expected for a volume bulk effect.
This shows that the above discussed low-order harmonic emission from the nanoplasma is a {\em surface} effect which becomes relatively less pronounced with increasing of the cluster size.
In small cold metal clusters subject to infrared laser pulses of moderate intensity the $R_0^4$ dependence was observed in \cite{lippitz} though the experiment was not entirely conclusive (see also \cite{zar06} for the respective theoretical analysis).

Figure~\ref{fig2} shows harmonic spectra $P(\omega)$ calculated from the above-described RSM.
Within the model, the dynamics and radiation of the electron cloud are governed mainly by the values of the two parameters $\omega_{\rm Mie}/\omega_l$ and $E_0/\omega_l^2R_0$, determining the possibility of resonant enhancements for particular harmonics and the number of excited harmonics in the spectrum.
The eigenfrequency of the electron cloud $\omega_{\rm eff}$ depends upon the amplitude of the oscillations (the maximum value ${\rm max}[\omega_{\rm eff}]=\sqrt{A_1}\omega_{\rm Mie}$ corresponds to small harmonic oscillations) and, therefore, upon the laser intensity.
As a result, at fixed cluster parameters the resonant enhancements appear at certain values of the intensity.
For the spectra of Fig.\ref{fig2}(a,d) and (c,f) the parameters were chosen such that the maximum value of the eigenfrequency exceeds the integers 3 and 5, respectively, so that, with increasing intensity resonant enhancements of the third and the fifth harmonic appear.
These enhancements are clearly seen from the comparison of these spectra with the ones calculated for the case shown in Fig.\ref{fig2}(b,e) when the maximum eigenfrequency ${\rm max}[\omega_{\rm eff}]\approx 4.4$ is far from both resonances.
In panels (a-c) the intensities are chosen such that the sphere's oscillation amplitudes are almost the same in each case so that the relative difference in the strength  of the third and the fifth harmonic (about three orders in magnitude) shows the efficiency of the resonant enhancement.
For the higher intensities in panels (d-f) these enhancements are even more pronounced.

A comparison between the  RSM and the PIC results shows that for the low-order harmonics the RSM provides a qualitatively correct description of the spectrum, including the effect of resonant enhancement.
This gives an additional justification of the RSM for applications to laser-driven nanoplasmas.
However, for relatively high harmonics (ninth or higher) the results predicted by the RSM appear to be qualitatively wrong.
Indeed, the RSM predicts the appearance of higher harmonics with increasing laser intensity, as it is seen from Fig.~\ref{fig2}(f) where the spectrum ends up at the 13-th harmonic.
By a moderate variation of the parameters inherent to the model, namely $\omega_{\rm Mie}/\omega$ and $a$, one may obtain even higher harmonics for the same intensities used for the spectra of Fig.~\ref{fig2}.
As expected for a classical system, no signature of a plateau emerges in the spectra.
However, the PIC simulations reported in \cite{kundu07} do not show any harmonics above the seventh within the studied domain of parameters, including the very high intensity of  $7.5\times 10^{17}$W/cm$^2$.
The same effect was found in \cite{antonsen05} where no harmonics above the ninth have been observed for intensities up to $10^{17}$W/cm$^2$.
In Ref.\cite{kundu07} we concluded from the inspection of individual PIC electron's trajectories that a dynamical instability induced by the resonant interaction of electrons with the time-dependent self-consistent field is responsible for the suppression of high harmonics.
Obviously, such a mechanism is beyond the RSM since the latter accounts for collective electron dynamics only.
In the next section we introduce a single-particle model suited for the analysis of the motion and the radiation of individual electrons so that the instability responsible for the suppression of high harmonics from cluster nanoplasmas under the above-described conditions can be studied.

\section{Suppression of high harmonics via stochastic behavior of individual electrons} \label{sohhvsboie}
It is known from numerical studies (see examples in \cite{fomytski04,antonsen05,last,brabec}) that in a laser-driven cluster the electron population separates into a dense core with a radius comparable to the initial cluster radius $R_0$ and a rarefied halo with a typical size of several $R_0$.
This subdivision is equivalent to a separation of quasifree electrons into deeply and weakly bound electrons, correspondingly.
A decrease of the electron density due to outer ionization is compensated by the inner ionization of atoms, provided inner ionization is not yet depleted, so that the cycle-averaged electron density distributions both in the core and the halo evolve rather slowly in time.
The core oscillates with relatively small deformations so that the RSM seems to be applicable to the deeply bound electron's dynamics while electron trajectories in the halo are strongly disturbed by the laser field and cannot be captured by the RSM.

In Ref.~\cite{kundu07} the radiation of individual PIC electrons has been considered (see Fig.~3 there). It was shown that electrons radiate harmonics as long as they move inside the dense core.
Being liberated from the core, electrons leave the cluster vicinity almost promptly, usually within a laser period.
This means, that the halo contains basically no permanent population, but consists almost entirely of electrons on their way out of the cluster.
Hence the electron density in the halo determines the rate of outer ionization from the cluster.
During the ejection, each electron emits an intense flash of radiation with an almost continuous spectrum that extends up to significantly higher frequencies than present in the net harmonic spectrum.
In Ref.\cite{kundu07} we argued that these flashes add up incoherently in the total emission amplitude.
Here we introduce a simple analytical 1D model which helps to illuminate the physical origin of this incoherency.
Despite its simplicity our model is able to describe, at least qualitatively, all essential features seen in the simulations.

\subsection{Model}
Let us suppose frozen ions and the number of nanoplasma electrons fixed so that in the absence of the laser field each electron moves in the time-independent self-consistent potential $U(x)$.
The laser field excites oscillations of the electron cloud which induce the ac part of the space charge field $E^{(I)}_{sc}(t)$.
The net oscillating field inside the system can be written as
\beq
{\cal E}(t)=E_l(t)+E^{(I)}_{sc}(t)\approx{\cal E}_0f(t)\cos(\omega_l t+\alpha),
\label{field}
\eeq
where $f(t)$ is the time-dependent pulse envelope.
If the Mie frequency notably exceeds the laser frequency, $\omega_{\rm Mie}>\omega_l$, as it is the case for infrared lasers, the laser field and the ac space-charge field $E^{(I)}_{sc}(t)$ essentially compensate each other.
In this case the amplitude of the net oscillating field inside the cluster ${\cal E}_0$ is related to the amplitude of the laser field $E_0$ according
\beq
{\cal E}_0\approx\frac{\omega^2_lE_0}{\omega^2_{\rm Mie}-\omega^2_l}\ll E_0.
\label{E0}
\eeq
This result is commonly referred to as {\em screening} of a low-frequency laser field inside clusters (see, e.g., the review \cite{krainov02}).
There is no contradiction between this screening of the laser field for individual electrons and the fact that the whole electron cloud feels the unscreened field.
Indeed, in the RSM there are two forces acting on the electron cloud: one due to the interaction with the ion core, another one due to the laser force, see Eq.(\ref{NE0}).
In the single-electron description we should also take into account the interaction between the electron under consideration and all other electrons in the cloud.
Within the model we assume that the electron cloud undergoes small, slightly nonlinear oscillations, so that this extra force is almost homogeneous inside the cluster and oscillates in time with the frequency $\omega_l$.
Within the RSM and under the condition $\omega_l\ll\omega_{\rm Mie}$ we assume throughout the paper that the electron cloud displacement $\vecx(t)$ reads
$$
\vecx(t)\approx-\frac{e{\cal E}_0}{m(\omega_{\rm Mie}^2-\omega^2)}f(t)\cos(\omega_lt+\alpha).
\nonumber
$$
Calculating the electric field induced inside the cluster due to this displacement and summing it up with the laser field (\ref{field}) one obtains the estimate (\ref{E0}) for the amplitude of the net oscillating field.
This type of screening results from the coherent superposition of the applied and the self-consistent field and has nothing in common with damping of electromagnetic waves in macroplasmas. The latter occurrs on  the spatial scale of the  skin-depth, which is in general much bigger than the typical cluster size we consider.

Taking a cluster consisting of $N\approx 1.7\times 10^4$ Ar atoms ($R_0\approx 6.2$nm) with the average ion charge ${\overline z}\approx 6$ and the degree of outer ionization $\eta\approx 0.5$, one can estimate $\hbar\omega_{\rm Mie}\approx 6$eV.
In a Ti:Sa laser pulse of intensity $5\times 10^{17}$W/cm$^2$ ($E_0\approx 3$a.u.) the amplitude of the oscillating field inside the cluster is according to (\ref{E0}) ${\cal E}_0\approx 0.2$a.u., i.e., more than one order of magnitude below the amplitude of the applied laser field.
The quasistatic part of the space-charge field $E_{sc}^{(II)}=E_{sc}-E_{sc}^{(I)}$
which traps electrons within the ion core can also be estimated for the assumed values of $\eta$ and ${\overline z}$.
Namely, the field near the cluster edge is $E^{(II)}_{sc}\approx\eta N{\overline z}e/R_0^2\approx 3$\,a.u., i.e., more than one order of magnitude above the oscillating field amplitude.
From this estimate we conclude that the oscillating field inside the cluster usually remains small compared to the quasistatic space-charge field.

Within the model the electron's evolution is governed by the Hamiltonian
\beq
H(p,x,t)=\frac{p^2}{2m}+U(x)-e{\cal E}(t)x\equiv H_0(p,x)-e{\cal E}(t)x
\label{H}
\eeq
and the corresponding Newton equation
\beq
{\dot p}=m{\ddot x}=-\frac{\partial U}{\partial x}+e{\cal E}(t)\equiv -eE^{(II)}_{sc}(x)+e{\cal E}(t),
\label{NE}
\eeq
where $m$ and $e$ are the electron mass and charge.
The well $U(x)$ is created by the quasistatic part of the space charge.
We model it by the function
\beq
U(x)=U_0[1-1/\sqrt{1+(x/R_0)^2}],
\label{c}
\eeq
where the values $R_0$ and $U_0$ are the cluster radius and the depth of the self-consistent well, respectively.
Here we choose the energy minimum $\epsilon=0$, so the $\epsilon=U_0$ is the continuum threshold.
For small excursion amplitudes this well is a nonlinear oscillator while for large excursions it has  the desired Coulomb behavior.
According to the estimates given above we assume the inequality
\beq
\mu\equiv\frac{e{\cal E}_0}{{\cal F}_0}\ll 1,~~~~~{\cal F}_0=\frac{U_0}{R_0}
\label{mu}
\eeq
being satisfied, where ${\cal F}_0$ has the meaning of a characteristic quasistatic force trapping the electron.

\subsection{Dynamics}
\begin{figure}
\includegraphics[width=0.3\textwidth,angle=-90.0]{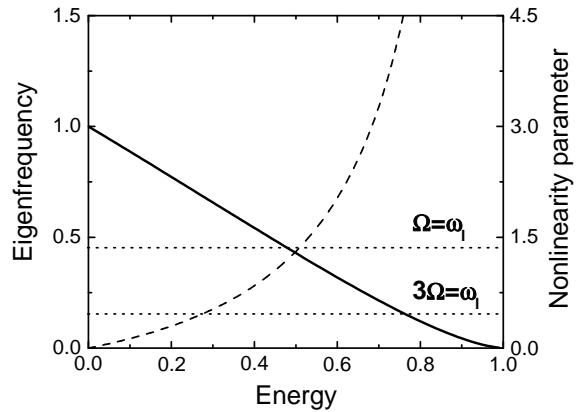}
\caption{Dynamical characteristics of the potential (\ref{c}): the scaled eigenfrequency $\Omega(\epsilon)/\Omega(0)$ (solid line) and the nonlinearity parameter (\ref{beta}) (dashed line) vs the scaled energy $\epsilon/U_0$. Positions of the first and third-order resonances are shown by dotted lines. The value of the parameter $\beta$ at the resonant energies is $\beta(\epsilon_1)\approx 1.2$ and $\beta(\epsilon_3)\approx 4.5$, respectively.}
\label{fig3}
\end{figure}

The dynamics of the unperturbed system with the Hamiltonian $H_0$ is characterized by the energy dependence of the eigenfrequency
\beq
\Omega(\epsilon)=\frac{2\pi}{T(\epsilon)},~~~~~
T(\epsilon)=\sqrt{2m}\int\limits_a^b\frac{dx}{\sqrt{\epsilon-U(x)}},
\label{T}
\eeq
where $T(\epsilon)$ is the oscillation period, $a(\epsilon)$, $b(\epsilon)$ are the turning points, and $\epsilon>0$ is the total energy \cite{goldstein}.
The energy-dependent parameter
\beq
\beta(\epsilon)=\frac{\epsilon}{\Omega(\epsilon)}\cdot\bigg\vert\frac{d\Omega}{d\epsilon}\bigg\vert
\label{beta}
\eeq
characterizes the nonlinearity of the unperturbed system and thus its potential capability to emit harmonics.
Figure~\ref{fig3} shows the energy dependence of the scaled eigenfrequency $\Omega(\epsilon)/\Omega(0)$ and the parameter (\ref{beta}).
In cluster potentials the period $T(\epsilon)$ increases with increasing energy so that
\beq
d\Omega/d\epsilon<0.
\label{domega}
\eeq
The well (\ref{c}) has this property.

\begin{figure}
\includegraphics[width=0.3\textwidth,angle=-90.0]{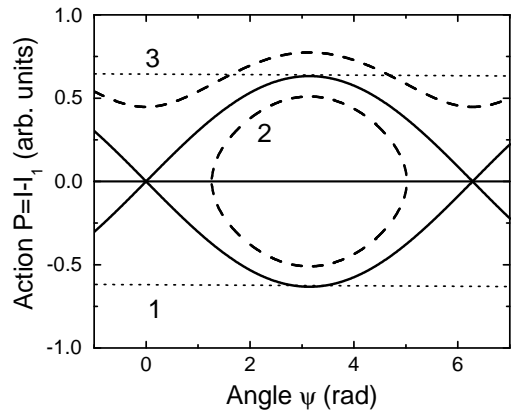}
\caption{The phase space of the Hamiltonian (\ref{Hfinal}). A separatrix (thick solid lines) subdivides the phase space into domains of finite (2) and infinite (1,3) motion. Two typical trajectories are shown by dashed lines. Dotted lines indicate the positions of the thresholds $\pm P_{\rm max}$ determined in (\ref{omph}).}
\label{fig4}
\end{figure}

The system (\ref{NE}) is not integrable.
However, if the time-dependent force remains small in comparison to the first term at the right hand side of (\ref{NE}) [i.e., under the condition (\ref{mu})] the related dynamics can be qualitatively described using general methods of nonlinear mechanics \cite{liber,sagdeev,chirikov}.
It is well established that the behavior of a nonlinear conservative system subject to a weak oscillating field is essentially governed by the resonances between the energy-dependent frequency of the unperturbed oscillations (\ref{T}) and the frequency $\omega_l$ of the external force.
The positions $\epsilon_{sm}$ of these resonances are determined by
\beq
s\Omega=m\omega_l
\label{rescond}
\eeq
with $m$ and $s$ integers.
In a harmonic oscillator only the resonance with $m=s=1$ can be excited.
Resonances with $m=2,3,...$ are nonlinear with respect to the external field ${\cal E}(t)$ (in a quantum picture such resonances correspond to multiphoton absorption and emission).
Resonances with $s=2,3,...$ are related to the nonlinearity inherent to the unperturbed system $H_0$ so that for a highly nonlinear system they may be significant even in the weak perturbation regime.
This situation corresponds to the case we consider so that below we assume $m=1$, disregarding the field-induced nonlinearities.
The reader should not be confused by the fact that we disregard here exactly the same nonlinearities which lead to the emission of harmonics we study in the rest of this work.
Although these weak nonlinearities are responsible for the emission they play no role for the stability analysis we consider in this subsection.
Because of inversion symmetry of the well (\ref{c}) only odd resonances with $s=1,3,5...$ are important.
Before the pulse an electron moves along an unperturbed trajectory, specified by the energy $\epsilon_0$ and the initial phase $\vartheta_0$.
If the energy is far from any resonance level $\epsilon_s$ and the condition (\ref{mu}) is satisfied the oscillating field in (\ref{NE}) induces a small perturbation and, as a consequence, the trajectory is only slightly disturbed by the field.
Under the condition (\ref{domega}) the first-order resonance $s=1$ is the lowest one on the energy scale so that the vast majority of such nonresonant trajectories lies well below the first resonant energy $\epsilon_1$.
This ``perturbative'' regime of interaction survives until the time-dependent energy approaches the vicinity of the resonance, either due to an increasing amplitude ${\cal E}_0$ or because of a higher initial energy of the particle.
Near the resonance the same small perturbation yields a strongly disturbed electron trajectory. 
A qualitative description of such trajectories and the determination of the boundaries that separate the  ``perturbative'' and the ``resonant'' domains is possible using  the methods described, e.g., in Refs.~\cite{liber,sagdeev,chirikov}.
We introduce new canonical variables, namely the  ``action'' $I$ and the ``angle'' $\vartheta$, which are defined in the standard way \cite{goldstein,sagdeev,liber}
\begin{equation}
I(\epsilon)=\frac{1}{2\pi}\oint p(\epsilon,x)dx,~~~~~
\vartheta=-\frac{\partial S}{\partial I}.
\label{ITHETA}
\end{equation}
Here $p=\sqrt{2m(\epsilon-U(x))}$ is the electron momentum in the unperturbed system and $S(x)$ is both the position-dependent reduced action and the generator of the canonical transformation.
The integration in (\ref{ITHETA}) is performed along the closed trajectory with the energy $\epsilon$.
Strictly speaking, this canonical transformation is defined for the unperturbed Hamiltonian $H_0$ only when closed, periodic trajectories exist.
In this case one obtains immediately the new canonical variable $\vartheta=\Omega(\epsilon)t+\vartheta_0$.
The variables $I,\vartheta$ provide an effective zero-order approximation for the construction of a specific perturbation theory capable of describing near-resonant dynamics \cite{sagdeev,chirikov,liber}.
Using the new variables and assuming that the particle energy is sufficiently close to a resonant value, say $\epsilon_1$, we may omit rapidly oscillating terms and obtain an approximate, time-independent resonant Hamiltonian
\beq
H_r(I,\psi)=H_0(I)-\omega_l I-\frac{e{\cal E}_0}{2}x_1(I)\cos(\psi),
\label{Hnew1}
\eeq
so that the corresponding system is integrable.
Here $x_1$ is the first Fourier-component of an unperturbed trajectory, ${\cal X}(I,\vartheta)=\sum_kx_k(I)\cos(k\vartheta)$ and $\psi=\vartheta-\omega_lt$.
Now the trajectories of the system in the new phase space can be found, at least in the form of integrals.
Further simplifications are possible if we take into account that for a weak perturbation the deviation of the action from its resonant value $I_1\equiv I(\epsilon_1)$ is small, i.e., $|I-I_1|\ll I_1$.
Introducing the new canonical variable $P=I-I_1$ we obtain the nonlinear oscillator Hamiltonian (with the negative effective mass):
$$
H^{\prime}_r(P,\psi)=-\frac{P^2}{2M}-B\cos(\psi),
\nonumber
$$
\beq
M=\frac{1}{\Omega_1|\Omega_1^{\prime}|}>0,
~~~~~B=\frac{e{\cal E}_0x_1(I_1)}{2}
\label{Hfinal}
\eeq
with $\Omega_1^{\prime}=d\Omega/d\epsilon|_{\epsilon=\epsilon_1}$.
The electron behavior can thus be qualitatively described as nonlinear oscillations in $(P,\psi)$ space, known as {\it phase oscillations} \cite{liber, sagdeev,chirikov}.
In the new canonical variables which have the formal status of momentum $(P)$ and coordinate $(\psi)$
the phase space of the Hamiltonian (\ref{Hfinal}) splits into domains of finite and infinite motion (see Fig.\ref{fig4}).
Finite motion corresponds to a particle trapped by the resonance, while particles moving infinitely do not intersect with the resonance.
The separatrix of (\ref{Hfinal}) is a boundary between these two domains.
The motion near separatrices is unstable, so that even a small variation in the initial conditions may entirely change a trajectory.
As a result, particles approaching a separatrix may penetrate from one domain to another or  may be trapped by a resonance.
The parameters characterizing the motion of trapped particles are the maximum deviation from the resonant action $I_1$ and the frequency of small phase oscillations:
\beq
P_{\rm max}=2\sqrt{MB}\simeq\frac{\epsilon_1}{\omega_l}\sqrt{\frac{\mu}{\beta}},~~~~~
\Omega_{\rm ph}=\sqrt{\frac{B}{M}}\simeq\omega_l\sqrt{\mu\beta}.
\label{omph}
\eeq
In energy space the positions of the separatrices of the first-order resonance are determined by
\beq
\epsilon_1^{\pm}=\epsilon(I_1\pm P_{\rm max})=\epsilon_1\pm \Delta\epsilon_1,~~~~~
\Delta\epsilon_1\simeq U_0\sqrt{\frac{x_1}{R_0}\frac{\mu}{\beta}}.
\label{threshold}
\eeq
If the energy intersects a respective threshold so that $\vert\epsilon-\epsilon_1\vert\le\Delta\epsilon_1$, the electron becomes trapped by this resonance domain and experiences phase oscillations with the frequency and amplitude both proportional to $\sqrt{\mu}$.
Due to the appearance of a new time scale given by the frequency of the phase oscillations (\ref{omph}) the motion becomes aperiodic and highly nonlinear.
It should be emphasized that in weak fields the perturbation parameter $\mu$ is far from resonance and $\sqrt{\mu}$ close to it so that the near-resonant motion appears to be much more perturbed than the off-resonant one.
\begin{figure*}
\includegraphics[width=0.5\textwidth,angle=-90.0]{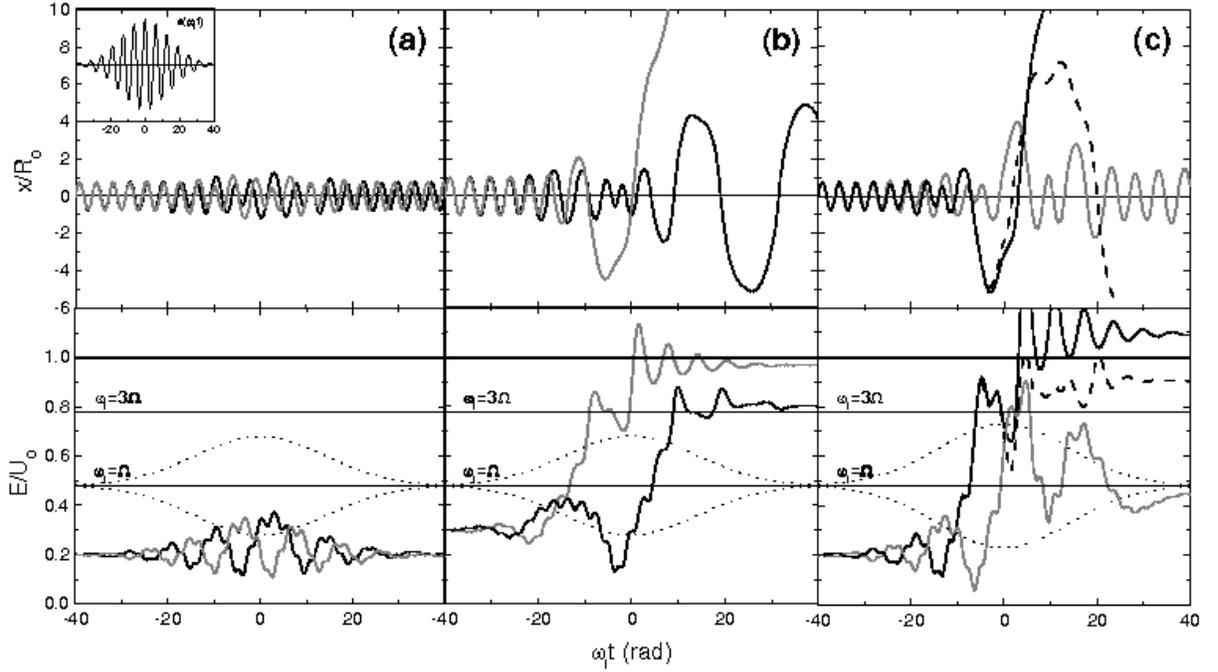}
\caption{Upper panel: electron coordinate vs dimensionless time, $x(\varphi=\omega_lt)$, calculated from the numerical solution of (\ref{NE}) for $U_0=5.0$, $\omega_l=1.0$, and $R_0=1.0$, assuming the initial conditions $x(-40)=0.0$, $0\le\alpha\le\pi/2$ (black solid and dashed curves) and $\alpha=\pi$ (gray curves). Other parameters are: $\epsilon_0=0.2U_0,~\alpha=0.0,~{\cal E}_0=0.5$ (a), $\epsilon_0=0.3U_0,~\alpha=1.0,~{\cal E}_0=0.5$ (b) and $\epsilon_0=0.2U_0,~{\cal E}_0=0.8$ (c) with $\alpha=0.0$ for a solid and $\alpha=0.05$ for a dashed curve. The electric field ${\cal E}(\omega_0t)$ is shown in the insert.
Lower panel: scaled electron energy vs dimensionless time for the parameters of the upper panel, a-c. Thin horizontal lines show the position of the first-order $(\epsilon_1\approx 0.48U_0)$ and the third-order $(\epsilon_3\approx 0.77U_0)$ resonances. Dotted lines show positions of the time-dependent separatrices (\ref{threshold}) for the first resonance. A thick horizontal line shows the continuum threshold $\epsilon=U_0$.}
\label{fig5}
\end{figure*}

The interaction with an isolated resonance cannot lead to ionization since the electron remains  trapped for an, in principle, infinite time \cite{comm3}.
However, the higher-order resonances lying above may come into play.
As soon as the separatrices of neighboring resonances intersect, the particle, captured by the first resonance, may jump to the third, etc.
Because the volume of the accessible phase space is increasing with energy this inter-resonance motion will have a predominant direction, namely towards higher energies.
This leads to a fast liberation of the electron from the system, known as {\em  stochastic ionization} \cite{chirikov,sagdeev,kost05}.
For realistic parameters of laser-cluster interaction this overlap of resonances is realized with almost 100\% probability so that cases where the particle remains trapped by a resonance are rare while almost prompt ionization occurs as soon as the first-order resonance is reached.

We visualize the above-described scenario by solving Eq.(\ref{NE}) numerically.
In the calculation we take $\omega_l=R_0=1.0$ and $U_0=5.0$ so that even the amplitude ${\cal E}_0=1.0$ ($\mu=0.2$) still corresponds to the weak-field regime, as defined above.
These parameters are not arbitrarily chosen.
Indeed, a solution of Eq.(\ref{NE}) with the well (\ref{c}) and the field (\ref{field}) with a slowly varying envelope depends on four dimensionless parameters, $\mu$, ${\cal F}_0/m\omega^2 R$, $\epsilon_0/U_0$ and $\alpha$, the two last of them defining the initial conditions.
To recalculate all the parameters for a real system we should assume some certain values for the cluster radius and the laser field frequency, which then define all other parameters.
For typical values, say $R_0=5$nm and $\hbar\omega_l=1.55$eV, and for the dimensionless parameters of Fig.~\ref{fig5} one may check that the resulting strengths of the quasistatic and the oscillating parts of the self-consistent field indeed correspond to the estimates given below Eq.(\ref{E0}).
The positions of the most important first and third-order resonances are $\epsilon_1\approx 0.48U_0$ and $\epsilon_3\approx 0.77U_0$, respectively.
A particle with the initial energy $\epsilon_0<\epsilon_1$ starts its motion at $x=0$ and $\varphi\equiv\omega_lt=-40$ when the field ${\cal E}(t)$ is negligibly small.
Then the field (with a Gaussian envelope) increases, and the electron propagation under the action of the full force is calculated until $\varphi=+40$.
By choosing different phases $\alpha$ of the field we model different initial conditions for the particle at the fixed initial energy $\epsilon_0$.
The results are summarized in Fig.\ref{fig5}.

Figure~\ref{fig5}a corresponds to the ``perturbative'' regime of interaction.
The initial energy is far enough from the first (lower) resonance, so that the trajectory in the energy space does not intersect the respective lower separatrix, or just touches it.
As a result, the trajectory remains weakly disturbed, its shape is well described as a superposition of oscillations with the frequencies $\Omega(\epsilon_0)$ and $\omega_l$.
By choosing different initial conditions we obtain trajectories simply shifted in time by the value of $\alpha$.
Figures~\ref{fig5}(b,c) correspond to the ``resonant'' regime of interaction where the initial energy is high enough or the field is strong enough to cause penetration of the particle into the vicinity of the first resonance.
The five trajectories plotted in Figs.\ref{fig5}(b,c) show that the near-resonant motion is very sensitive both to the initial conditions and to the field amplitude so that a particular trajectory appears to be unpredictable.
Usually the particle is emitted from the system while in rare cases it remains in a bound state after the pulse is off, being trapped by a resonance (see the trajectories in Fig.~\ref{fig5}(b,c)).
From this observations we may conclude that at parameters typical for intense laser-nanoplasma interactions the particle behavior in the ``resonant'' regime becomes stochastic, as it is expected to be according to the general theory \cite{liber, sagdeev,chirikov}.

\begin{figure}
\includegraphics[width=0.3\textwidth,angle=-90.0]{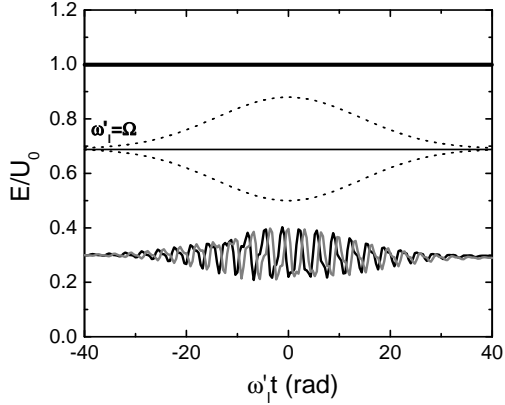}
\caption{Same as the lower panel of Fig.~\ref{fig5}b but for half the frequency, $\omega_l^{\prime}=0.5$ and the duration two times longer, so that the dimensionless time $\varphi=\omega_l^{\prime}t$ varies within the same limits.}
\label{fig6}
\end{figure}

In classical ionization, a particle has to overcome the potential barrier, i.e., its total energy must exceed the maximum value of the potential energy suppressed by the field at some time instant.
%Although this definition of the ionization threshold is strictly valid in the quasi-static limit of low laser frequencies, calculations of trajectories show that it remains relevant in the case we consider $\Omega(\epsilon_0)\simeq\omega_l$ too.
Obviously, the trajectories of Fig.~\ref{fig5}(b,c) satisfy this condition, while both trajectories of Fig.~\ref{fig5}a do not.
%With this definition, ionization and excitation can be explained without appealing to the resonant picture.
%The field amplitude ${\cal E}_0$ in Fig.~\ref{fig5}(b,c) is already close to the value for which the initial electron energies $\epsilon_0=0.3U_0$ and $\epsilon_0=0.2U_0$ are just slightly below the maxima of the potential barrier suppressed by the field (\ref{field}), $0.38U_0$ and $0.23U_0$, respectively.
%Thus the question may arise, whether the above-described stochasticity followed by ionization or excitation indeed results from a resonant interaction.
In Fig.~\ref{fig6} we show the trajectories in energy space evaluated for half the laser frequency, $\omega_l^{\prime}=\omega_l/2=0.5$, and all other parameters the same as in Fig.~\ref{fig5}b.
The plots show that with decreasing laser frequency the time-dependent energy gain from the field to the particle decreases down to a perturbative level, the total energy remains always below the maximum of the time-dependent potential energy, and no ionization or excitation occurs. 
Within the resonant picture we exploit here the qualitative difference between the trajectories of Figs.~\ref{fig5}(b,c) and \ref{fig6} appears because with decreasing frequency the first resonant level is shifted up and no penetration into the resonant area between the separatrices takes place anymore.
%If, on the other hand, the resonant interaction is not significant the particles should escape more easily for the lower frequency because in this case the barrier is suppressed for a longer time and the particle has more time to be accelerated by the field.

It is instructive to show the connection between the single-electron model specified by Eqs.~(\ref{field}),(\ref{H}),(\ref{NE}),(\ref{c}) and a model which describes collective motion, as the RSM of Sec.~\ref{reoloh} does.
The RSM deals with the electron cloud displacement $\vecx(t)$ whose Fourier-transform is directly related to the spectrum.
This value can also be calculated within a single-electron picture as
\beq
\vecx(t)=\int d\epsilon_0\int d\alpha\ \vecr(\epsilon_0,\alpha,t)F(\epsilon_0,\alpha),
\label{Xt}
\eeq
where $\vecr(\epsilon_0,\alpha,t)$ is the individual trajectory with the initial energy $\epsilon_0$ and the initial condition $\alpha$, and $F(\epsilon_0,\alpha)$ is the distribution function for electrons before the field is on.
One should note that, contrary to the RSM, the spatial distribution of the electrons in the presence of the field is not known unless one calculates all individual trajectories.
% so that the total force due to electron-ion interaction can not be straightforwardly found without calculating the individual trajectories.
Calculating a trajectory $\vecr(\epsilon_0,\alpha,t)$ analytically is possible only within perturbation theory with respect to the external field where one may easily derive (\ref{NE0}) with the linear part of the restoring force only from (\ref{Xt}).
The derivation of nonlinear corrections, although doable, requires very cumbersome algebra.
The single-particle model is appropriate for a qualitative description of the stochastic resonant behavior but hardly applicable to the study of the slightly unharmonic motion of the deeply bound electrons.

\subsection{Radiation}
\begin{figure}
\includegraphics[width=0.4\textwidth]{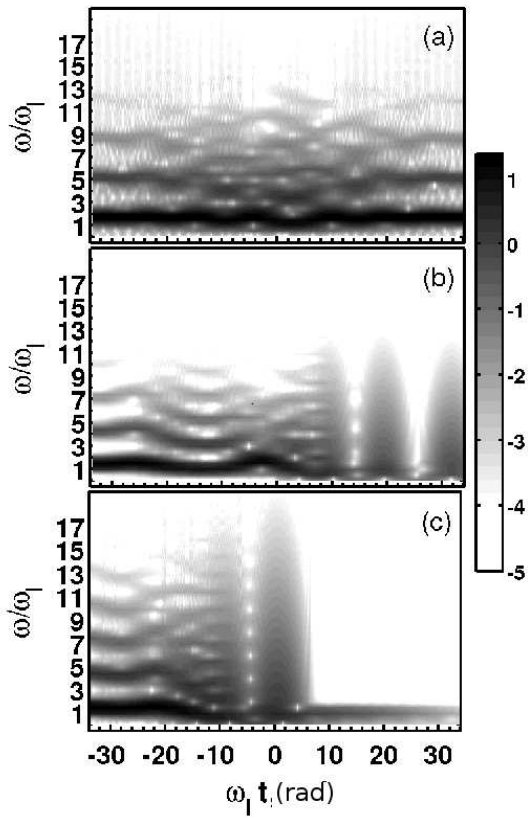}
\caption{TF diagrams corresponding to three (the black curve of Fig.~\ref{fig5}a and both curves of Fig.~\ref{fig5}b) electron trajectories.}
\label{fig7}
\end{figure}

The analysis of the previous subsection gives a direct explanation of the radiation spectra extracted from the PIC results in Ref.~\cite{kundu07} both for individual electrons and for whole clusters.
Deeply bound electrons with energies below the first resonant separatrix $\epsilon\le\epsilon_1-\Delta\epsilon_1$ move along slightly perturbed regular trajectories.
This causes HG with rapidly decreasing yield as a function of the harmonic order so that even the seventh harmonic is barely present in the corresponding spectrum of Fig.~\ref{fig7}a.
An individual electron, while passing the resonance and being trapped by it or leaving the cluster potential, emits radiation due to its strong acceleration, seen as a flash in the TF spectrograms of Fig.~\ref{fig7}b,c.
These spectrograms should be compared with the ones extracted from our PIC results (see Fig.3 in \cite{kundu07}).
Exactly because of the stochastic nature of nonlinear resonance the electrons' trajectories are very sensitive to the initial conditions with which the nonlinear resonance is entered, as is clearly seen from Fig.~\ref{fig5}c where solid black, dashed black and gray trajectories correspond to the same initial energy and the same field amplitude but different phases of the external field.
As a result, flashes from different electrons are incoherent (the corresponding amplitudes have nearly random phases), and, being added up, vanish in the total dipole acceleration.

This shows that exactly the same mechanism behind efficient energy absorption by  and outer ionization from clusters, namely nonlinear resonance \cite{mulser05,kost05,antonsen05,kundu06}, restricts HG from them by breaking the coherent electron motion once it becomes strongly anharmonic.
Only well-bound electrons trapped inside the ionic core with energies far from the resonance contribute to the net, coherent radiation of the cluster.
A similar behavior was observed in classical ensemble simulations of atomic HG \cite{bandarage}.

Instabilities induced by the resonant interaction grow in time (usually with an exponential rate).
Thus the picture depends also on the pulse duration.
With all other parameters fixed, for longer pulses, more and more trajectories from the vicinity of the separatrix experience a stochastic behavior.
As a consequence, an increase of the pulse duration should lead, in general, to a  further loss of coherency.

\section{Conclusions}
Although laser-irradiated cluster nanoplasmas emit low-order harmonics efficiently, no significant yield of high harmonics can be expected even for very high laser intensities.
This is  a consequence of dynamical stochasticity, inherent to nonlinear dynamical systems driven by weak, time-dependent forces such as the screened electric field inside a cluster.

Increasing the laser intensity does not help much because the self-consistent field trapping the electrons inside the ion core increases too. As a consequence the  electron population always splits into deeply bound electrons and a halo, the Mie-frequency and the screening increases, so that the physical picture remains almost insensitive to the intensity of the applied field.

Another option we did not consider above is to use relatively long pulses where the first-order resonance with the Mie frequency can be reached because of the ion core expansion.
If the Mie resonance is met the ac electric field inside the cluster is grossly enhanced and nearly all electron trajectories appear to be strongly disturbed.
In this case an analysis of near-resonant stochastic behavior as given above is inappropriate.
It seems that a direct numerical study is the only option under these conditions.
An analysis based on classical Vlasov simulations was performed in Refs.[14].

\bigskip

\section{Acknowledgment}
We are grateful to W.\ Becker for valuable discussions.
This work was supported by the Deutsche Forschungsgemeinschaft and the Russian Foundation for Basic Research (project No.06-02-04006).


\begin{thebibliography}{99}
\bibitem{comm3} Unless the amplitude of phase oscillations around the first resonance is already enough to lift the electron above the continuum threshold.
\end{thebibliography}
\end{document}